\documentstyle[12pt]{article}

\begin{document}
\begin{titlepage}
\begin{center}

June 28, 2000     \hfill    LBNL-43201 \\

\vskip .5in

{\large \bf Nonlocality, Counterfactuals, and Consistent Histories}
\footnote{This work was supported in part by the Director, Office of Science, 
Office of High Energy and Nuclear Physics, of the U.S. Department of Energy 
under Contract DE-AC03-76SF00098.}
\vskip .50in
Henry P. Stapp\\
{\em Lawrence Berkeley National Laboratory\\
      University of California\\
    Berkeley, California 94720}
\end{center}

\vskip .5in

\begin{abstract}
Numerous experiments pertaining to quantum nonlocality are underway
or have recently been completed. A certain theoretical defect in the linkage
between these experiments and nonlocality is discussed. This defect consists 
of the use of a reality concept that conflicts with the quantum precepts. 
This concept could be a hidden-variable assumption, or an Einstein reality
assumption. A proof that avoids these difficulties has been given recently. 
It demonstrates the incompatibility of the predictions of quantum theory 
with a certain locality requirement. This proof is based directly on 
logical arguments involving counterfactual statements, but has been 
criticized because it is based on orthodox logical principles, which 
might possibly be tainted with some classical idea. A rigorous framework 
for reasoning with counterfactual statements within the structure defined 
by the quantum precepts is constructed here, and the earlier proof is carried 
through in this strictly quantum framework. Griffiths has proposed another 
framework for consistent reasoning with counterfactuals within quantum theory,
and the proof goes through also within the Griffiths framework. The proof 
involves a Hardy-type of experimental set up, and, thus, according to the 
present analysis, it is experiments of this kind that should be performed to 
validate the empirical basis of any claimed quantum nonlocality.

\end{abstract}
\medskip
\end{titlepage}

\renewcommand{\thepage}{\roman{page}}
\setcounter{page}{2}
\mbox{ }

\vskip 1in

\begin{center}
{\bf Disclaimer}
\end{center}

\vskip .2in

\begin{scriptsize}
\begin{quotation}
This document was prepared as an account of work sponsored by the United
States Government. While this document is believed to contain correct 
 information, neither the United States Government nor any agency
thereof, nor The Regents of the University of California, nor any of their
employees, makes any warranty, express or implied, or assumes any legal
liability or responsibility for the accuracy, completeness, or usefulness
of any information, apparatus, product, or process disclosed, or represents
that its use would not infringe privately owned rights.  Reference herein
to any specific commercial products process, or service by its trade name,
trademark, manufacturer, or otherwise, does not necessarily constitute or
imply its endorsement, recommendation, or favoring by the United States
Government or any agency thereof, or The Regents of the University of
California.  The views and opinions of authors expressed herein do not
necessarily state or reflect those of the United States Government or any
agency thereof or The Regents of the University of California and shall
not be used for advertising or product endorsement purposes.
\end{quotation}
\end{scriptsize}

\vskip 2in

\begin{center}
\begin{small}
{\it Lawrence Berkeley Laboratory is an equal opportunity employer.}
\end{small}
\end{center}

\newpage
\renewcommand{\thepage}{\arabic{page}}
\setcounter{page}{1}

{\bf 1. Introduction.}
\vspace{.1in}

There is considerable experimental activity on the issue 
of quantum nonlocality. A recent issue Physics Today[1] 
has a bulletin entitled  ``Nonlocality Get More Real'' that reports 
experiments at three laboratories (Geneva, Innsbruck, and 
Los Alamos) directed at closing loopholes in proofs purporting to show
that ``seemingly instantaneous'' influences can act over large distances.
The first of these papers [2], where the measured effect extends over a 
distance of more than 10km, begins with the words  ``Quantum theory is 
nonlocal.''  The longer version [3]  says ``Today, most physicists 
are convinced that a future loophole-free test will definitely demonstrate 
that nature is indeed nonlocal.''

The question thus arises: Is there evidence for a breakdown of the
orthodox basic locality property of quantum theory?

The answer is ``No''.

But then what are these laboratories and physicists doing?

The answer is this: They are  considering a locality property that 
is related to the orthodox one, but not identical to it. But
what motivates this deviation from orthodoxy?

To understand the motivation one should recall first the orthodox locality
property. It follows directly from  basic principles of quantum theory.
These basic principles are:\\

1. The Reduction Formula:\\
\hspace*{1.1in}$ S \longrightarrow [ PSP + (1 - P)S(1 - P)$]\\
\hspace*{1.3in}$\longrightarrow [PSP$   or $(1 - P)S(1-P)]$.\\

2. The Probability Formula:\\
\hspace*{1.0in}$<P> = $ Trace $PS$/ Trace $S$.\\

3. The Microcausality Condition:\\
\hspace*{.9in}$ [Q_1(x_1), Q_2(x_2)]  = 0$ \hspace{.5in} for  
$(x_1 - x_2)^2 < 0$.  
\newpage
The first line of the first formula  specifies the reduction of the state $S$
associated with the answering (by nature) of the question associated with 
the projection operator with $P$,  provided that answer is {\it not} known: 
the second line describes the reduction when the answer {\it is} known.

The second formula is the prediction of quantum theory for the 
average value of the function that has value unity or zero according to
whether the answer to the question associated with the projection operator
P is `Yes' or `No', respectively.

The third line asserts that the operators associated with
observables measured in two space-like-separated regions commute.

If two projection operators $P_1$ and $P_2$  correspond to observables
measured in two spacelike-separated regions then these principles entail 
(using the normalization Trace S = 1 and the defining property of projection 
operators, $P^2 = P$)\\

\hspace*{.3in} $<P_1>'  = $  Trace $P_1 [P_2  S  P_2 + (1-P_2)S(1-P_2)]$\\
\hspace*{1.2in}=  Trace $P_1 [  S  P_2 P_2 + S(1-P_2)(1-P_2)]$\\
\hspace*{1.2in}=  Trace $P_1 [  S  P_2 + S(1-P_2)]$\\
\hspace*{1.2in}=  Trace $P_1 S$ \\
\hspace*{1.2in}= $ <P_1>$.\\

\noindent This gives the orthodox locality property:\\
 
\noindent ``The fraction of answers `Yes'  predicted by quantum theory for 
the outcome of a measurement performed in one spacetime region is 
independent of which experiment, if any, is performed in a spacetime 
region that is space-like-separated from the first region.''
  \\

Each prediction of quantum theory is fundamentally a prediction of the
observed average value of a function that is either unity or zero depending
on whether some particular outcome appears or not. But the fact that such
an {\it average value} does not depend upon which experiment is performed
in a far-away region does not entail that the {\it individual
outcomes} do not depend upon which experiment is performed faraway: the
individual outcomes in one region could be highly dependent upon which
experiment is performed far away without violating any principle
of quantum theory.  Yet in spite of the silence of quantum
theory on this matter, it would nevertheless  seem that a dependence of the 
observed outcome in one region upon which free choice is made faraway at 
the same  time would violate {\it in some way} the idea that there is no 
faster-than-light influence of any kind. 

What will be proved here is this: The predictions of quantum theory are 
incompatible with the demand that no effect of any kind can be located 
outside the forward lightcone of its cause. This condition will be formulated
below as two precisely defined conditions, LOC1 and LOC2.
 
According to orthodox quantum thinking,  experimenters are
deemed free to choose which experiment they will perform. Indeed, Bohr's
idea of complementarity is that, for any single experiment that an 
experimenter might choose to perform, the quantum state of a system gives
statistical information pertaining to the outcomes that will appear to him 
if he chooses to perform that particular experiment. 

This idea, that the choices made by the experimenters can be treated
as free variables, underlies the idea of locality described above: it allows 
these choices to be identified as effective primitive causes, within the
context of the study of the system upon which the experiments are performed.

The three papers [2, 4, 5] cited in [1] do not actually examine any
locality property itself. They explore something else, namely whether
certain ``Bell Inequalities'' hold. John Bell [6], in a response to the paper
of Einstein, Podolsky, and Rosen [7] introduced a `reality' concept.
He {\it assumed}, in connection with certain two-particle correlation 
experiments, the existence of a ``hidden variable'' substructure, and 
then imposed on this substructure a locality condition. He showed
that the resulting local hidden-variable structure is incompatible with
the assumed validity of certain predictions of quantum theory. However,
the local hidden-variable structure that he assumed entails [8,9] the 
existence of a model in which the outcomes of all of the possible 
experiments that might be performed on the two particles are 
{\it simultaneously well defined}. Thus this assumed property directly 
violates a basic precept of quantum thinking.

Bell's objective was to show that the existence of an underlying reality 
of the sort that Einstein was searching for is incompatible with the 
{\it predictions} of quantum theory. Bell's success constitutes 
strong support for the orthodox quantum position that no underlying 
reality compatible with the precepts of classical physics can be compatible 
with the predictions of quantum theory. But since conclusions of this 
Bell-type depend upon assuming a hidden variables substructure (or making an 
associated ``Einstein Reality Assumption'') they do not entail that locality 
itself fails. For the hidden-variable (or Einstein Reality) assumption could 
be the source of the contradiction.
   
The basic problem that must be solved if one is to examine the
locality issue itself is this: How can one introduce a condition that
implements the locality notion --- that earlier effects cannot depend 
upon a later free choice between mutually exclusive experimental 
arrangements --- without introducing some structure that already 
violates the quantum precepts.  

The contradiction at issue here is between a locality condition and certain 
predictions of quantum theory. All of the relevant conditions can be 
formulated as logical statements. Hence the most direct approach would be 
to deduce rigorously a logical contradiction from the principles of logic 
alone, thereby avoiding any dependence of the conclusion upon other 
assumptions.

Logicians have developed  methods for reasoning consistently with 
statements of the relevant kind. Those methods provide a potent 
logical framework in which the arguments can be pursued. However, 
those orthodox principles of reasoning were developed in a context in which 
the notions of classical mechanics were tacitly presumed to be valid. So the 
question arises as to whether use of orthodox logic would already prejudice 
the argument.

It turns out that the orthodox principles work even better in 
the quantum context than in the classical one. This is because in the 
classical case the idea of a free choice between alternative possibilities 
already conflicts with the deterministic laws of physics that constitute
the foundation of the claim that some statement ``would be true'' if some 
choice had gone the other way. Moreover, in the classical context it 
was necessary to introduce an imprecise idea of ``closeness of worlds'' 
that takes one outside what is strictly deducible from the laws and 
principles of physics alone. In the present case the conclusion must 
be rigorous consequences of nothing but (1), the predictions of quantum 
theory; (2), a precisely formulated principle that expresses nothing but 
the locality concept in question; and (3), the rudimentary principles of 
logic. Any reliance on some vaguely defined concept of closeness is 
unacceptable.

In an earlier description [10] of the logical argument in question I 
stressed concordance with orthodox logic developed by logicians. 
However, that approach incurred a liability as regards communication with 
physicists. Most physicists are not versed in these logical developments, 
and lack the time to become so. Hence the contradiction could be imagined 
to be due to some spurious effect of an obscure logical framework. Adherence
to the orthodox principles was thus transformed from a virtue to a defect.

To avoid these problems I shall give here an internally complete account 
based exclusively on the precepts of quantum physics and 
rudimentary logic.

The question arises as to whether this issue is pertinent to physics.
One view would be that we have, in quantum theory, an excellent theory 
that appears to work perfectly, and that we should therefore rigorously 
abstain from thinking about questions that lead beyond the computational rules 
themselves: physicists should follow the adage ``Don't think, just calculate!'' 

That is a good recipe for developing in more detail what we already have.  
Yet physics is not engineering. The question of how the world can 
possibly be like what quantum theory shows it to be is not {\it necessarily} 
eternally unanswerable. Bell himself argued that we physicists ought to try 
to do better than just settle for puzzling rules about what will appear 
under specified conditions. Certainly the large efforts that 
experimentalist are putting into these probes into ``local realism'' 
and ``nonlocality'' show that many physicists are interested in delving 
into the odd behaviour of nature that is so wonderfully 
codified in the rules of quantum theory. Understanding the locality 
issue is one place where progress may be possible.

This matter has been discussed twice before in this journal.  

Unruh [11] has challenged my claim that the  method I employ 
avoids any violation of the quantum precepts.  His claims are based,
I believe, on a misapprehension of the structure of my argument.
Indeed, he proposed a wide assortment of conceivable interpretations of 
statements my in proof, and complained that I did not explain my approach in 
enough detail. So my first aim here is to describe the method in more detail
than before, completely within the framework provided by quantum precepts.

Griffiths [12] has proposed a quantum theoretical framework for consistent 
counterfactual reasoning based on his idea of ``consistent histories''.  
He applied his framework in one special way to one particular nonlocality
argument, of his own making, and found that this argument did not go through, 
and concluded, more generally, that there was no evidence within his framework 
for any nonlocal influences. However, Griffiths' framework for consistent 
counterfactual reasoning within quantum theory can be applied directly to the 
first part of my nonlocality argument, and  it then validates 
straightforwardly, within that framework, this first part of my argument. 
Applied next to the rest of my argument his framework validates the further 
claim that imposition of the second locality condition, LOC2, generates a 
contradiction. Thus Griffiths' framework for consistent counterfactual 
reasoning within quantum theory reaffirms the validity of the conclusion 
deduced here.

\vspace{.2in}
{\bf 2. Formulation of the First Locality Condition.}
\vspace{.2in}

The first locality condition expresses the locality/causality idea that 
if an experiment is performed and the outcome is recorded $\it prior$ to 
some time $T$, as measured in some Lorentz frame, then this outcome can be 
regarded as fixed and settled, independently of which experiment may be freely
chosen and performed (faraway) at a time later than $T$.

This putative condition is a theoretical idea, and it depends on another
theoretical idea, namely the notion that experimenters can be considered free
to choose between the  alternative possible experiments that are available 
to them. Bohr himself often stressed that the choices made by experimenters 
should be considered free: the whole idea of complementarity is that the 
single quantum state represents, simultaneously, the pertinent information 
concerning $\it all$ of these alternative possibilities. In his debate with
Einstein he never tried to duck the issues by claiming that one simply could 
not even contemplate or discuss these alternative possibilities, only one 
of which could actually be realized. By simply refusing even to contemplate 
the alternative possible experimental choices  Bohr could have protected 
quantum theory from challenges pertaining to possible nonlocal influences, 
but only at the expense of a closed mindedness that he did not embrace.

It turns out that one can, in fact, accommodate, without any apparent 
contradiction, the theoretical notion that recorded outcomes are
independent of what experimenters decide to do later. This can be achieved
by bringing counterfactual notions into quantum theory in a certain 
prescribed way.

The basic notion underlying a clean and rigorous approach to counterfactuals
within quantum theory is the concept of ``possible worlds''. Suppose one is 
considering a specified set of alternative possible experimental arrangements 
generated by the free choices made by a set of experimenters. Each of the 
alternative possible experimental arrangements is assumed to be specified 
by a set of local macroscopic conditions --- one for each experimenter --- 
localized in a corresponding spacetime region. And each of the possible 
outcomes of such an extended (global) experiment is supposed to be 
specified by a set of local macroscopic conditions, one located in each of 
the experimental regions. If there were $N_E$ experimenters in $N_E$ 
different regions, and each experimenter could choose between $N_M$ local 
measurements, and each of these measurements could have $N_O$ possible 
outcomes, then the total number of {\it logically possible} worlds under 
consideration is $(N_M \times N_O)^{N_E}$. Each of the finite set of 
elementary statements that enter into the characterization of these various 
worlds is associated with a macroscopic spacetime region, and with one bit of 
information that is associated with some possible macroscopic event in that 
region. The {\it physically possible} worlds are a subset of the logically 
possible worlds: the physically possible worlds include only those that, 
according to the predictions of quantum theory, have a non-null probability 
to appear. The physically possible worlds are called ``possible worlds.'' 
Normally, I omit also the word ``possible'': unless otherwise stated a 
``world'' will mean a ``physically possible world''.

The rudimentary logical relationships involve the terms ``and'', 
``or'', ``equal'' and ``negation''. A statement S involving these relations
is said to be true at (or in) world W if and only if S is true by virtue
of the set of truths that define W.

One further rudimentary relationship is the so-called ``material 
conditional'', which is represented here by the single arrow $\rightarrow$: 
the statement ``$ A \rightarrow B $ is true at world W'' is equivalent to 
[`A is false at W ' or `B is true at W ' ]. 

This rudimentary relationship is different from the logical relationship 
called the ``strict conditional'', which is represented here by the word 
``implies''. The statement `` `A is true' implies `B is true' '' is 
sometimes shortened to ``A implies B'', and is represented symbolically 
here by A $\Rightarrow$ B. By definition, $A\Rightarrow B$ is true if and 
only if for {\it every} (physically possible) world W either ``B is true 
at W'' or ``A is false at W'': i.e.,  for {\it every} (physically possible)
world W, the rudimentary statement $A\rightarrow B$ is true at W. 

The logical structure being used here can be expressed in terms of {\it sets}:
Let $\{W:X\}$ represent the set of worlds $W$ such that the rudimentary
statement $X$ is true at $W$. The symbol $\{X\}$ is an 
abbreviation of $\{W:X\}$.  Thus the statement that $A\Rightarrow B$ is true
is equivalent to the statement that $\{A\} \subset \{B\}$: [`The set of W at 
which A is true' is a subset of the set of W at which B is true.]
It is also equivalent to the statement: $\{A\}\cap\{\neg B\} = \emptyset$:
[The intersection of the set of W at which A is true with the set of W at
which B is false is the empty set.]

It follows from these definitions (see Appendix) that
$$
 [A \Rightarrow ( B \rightarrow C)]
 \equiv [(A \wedge B) \Rightarrow C], \eqno(2.1)
$$
where the symbol $\wedge$ stands for ``and'' (conjunction): Each side of (2.1)
is true if and only if $\{A\}\cap\{B\}\cap\{\neg C\}=\emptyset$.

Consider, then, the statement
$$
A \Rightarrow ( B \rightarrow C). \eqno(2.2) 
$$
But suppose  A is the negation of B: $A =\neg B$.  Then the statement (2.2)
is, by virtue of the identity (2.1), and the falseness of
``$\neg B \wedge B$'', true for any  C: the statement (2.2) would be true, 
but could have no empirical content. The same lack of content would
obtain if $A = (\neg B \wedge D)$.

Consider, then, the following analogous statement:\\
  \\
 ``If experiment R2 is performed in region R and experiment L2 is 
performed in (the spacelike separated) region L and outcome L2+ appears
in L, then if R1 is performed in R the outcome in L would  be L2+.''\\
  \\
Suppose  R1 and R2 are mutually exclusive experiments [i.e., 
``R2 is performed'' implies that it is false that ``R1 is performed'', 
and vice versa]. Then the above statement is trivially true for any 
value of the {\it final} symbol L2+. Consequently, this statement cannot 
be a valid expression of any locality property.\\

The correct statement of the locality condition employs a different sort 
implication, one that uses the word and concept ``instead''.  This concept
is of central importance in orthodox counterfactual reasoning, and I shall
here give it a precise and appropriate meaning within a strictly quantum 
context.

Consider a statement of the form:\\
  \\
LOC1: ``A implies that if, instead, C then D,'' \hfill (LOC1a)\\
  \\
where C might be incompatible with A. The premise of the final conclusion D  
is that C holds {\it instead} of A, not {\it in addition} to A. 
This avoids the intrinsic contradiction that arose before.
However, the exact meaning of statements of this form must be specified. 

The key statement ``If, instead, $C$ then $D$'' is traditionally represented 
symbolically by $[C \Box \rightarrow D]. $  I shall use that symbolic form 
for the quantum version defined here.  It is a statement that is made 
in one world, say $W$, about some other  worlds $W'$. I shall use it
to give precise meaning to the locality idea that no influence can 
have an effect outside the forward light cone of its cause.  However, 
the definition will entail all of the properties of these ``instead''
statements that arise in my proof, and that in the proof of ref. [10] were 
justified by appeal to orthodox logic. The condition $C$ in such a statement
will always be a condition that is controlled by a free choice: it will be an 
assertion that, in some one of the specified experimental spacetime regions, 
some one of the specified alternative possible mutually exclusive 
experiments associated with that region is chosen and performed.

The assertion $[C\Box \rightarrow D]$ is, by definition, true in world 
$W$ if and only if $D$ is true in {\it every} world $W'$ that is the same 
as $W$ apart from effects that could be due to imposing condition $C$. 
All effects are permitted except those that violate the locality condition 
or the predictions of quantum theory. This locality condition specifies, 
precisely, that all effects of imposing condition $C$ on the world $W$ are 
confined to the forward lightcone  $V^+(C/W)$ of the region in which $C$ 
conflicts with $W$. 

Thus $[C\Box \rightarrow D]$ is, by definition, true in $W$ if and only if
$D$ is true in every (physically possible) world $W'$ such that;\\
(1) $C$ is true in $W'$, and                        \hfill (2.5a)\\
(2) $W'$ coincides with $W$ outside $V^+(C/W)$.     \hfill (2.5b)   

The statement $[A \Rightarrow (C \Box \rightarrow D)]$ is, by definition,
true if and only if the set of possible worlds in which A is true is
contained in the set of possible worlds in which $[C \Box \rightarrow D]$
is true: implication retains its usual meaning.

The statement

LOC1: A $\Rightarrow$ [C $\Box \rightarrow$ D] \hfill (LOC1b)\\
captures exactly the locality idea that if $A$ is true in a possible world $W$
then $D$ must be true in every possible world $W'$ that differs from $W$ 
only by possible effects of imposing condition $C$. The only limitation
in these possible effects is that they are, by virtue of our locality 
condition, confined to the forward light-cone $V^+(C/W)$. 

An important special case is
$$
(L2\wedge R2\wedge L2+)\Rightarrow [R1\Box\rightarrow L2\wedge R1
\wedge L2+]. \eqno(LOC1c)
$$
This asserts that what is true in L, namely that the outcome + that (in 
some Lorentz frame) has already appeared, cannot be disturbed by a change in 
what will be freely chosen and performed at some later time. 

The definition of LOC1 entails the following property: If condition $B$
is located outside $V^+(C)$, hence outside $V^+(C/W)$ for all $W$, then
$$
[(A\wedge B)\Rightarrow (C\Box\rightarrow D)]\\
\equiv [A\Rightarrow (C\Box\rightarrow (B\rightarrow D))]. \eqno (LOC1d)  
$$
This follows from the fact that on the LHS of (LOC1d) the condition $B$ is 
imposed on $W$, hence, by virtue of condition (2.5b), on $W'$, whereas
on the RHS of (LOC1d) this condition $B$ is imposed on $W'$, hence, by virtue 
of condition (2.5b), on $W$.
 
A similar argument gives, under the same condition on $B$,
$$
[(A\wedge B)\Rightarrow (C\Box\rightarrow B\wedge D)]\\
\equiv [(A\wedge B)\Rightarrow (C\Box\rightarrow D)]. 
\eqno (LOC1e)  
$$
It also follows from the definition of LOC1, under this same condition on $B$.
that
$$
[B\Rightarrow (C\rightarrow D)]\Rightarrow 
[B\Rightarrow (C\Box\rightarrow D)]. \eqno (LOC1f)
$$

Proofs of these properties are given in the appendix.

\vspace{.2in}
\noindent{\bf 3.  Proof of the incompatibility of locality
with the predictions of quantum theory.}
\vspace{.2in}

The argument is based on a Hardy-type [13] experimental set-up. This set-up
defines the universe of statements under consideration here. 

There are two experimental regions $R$ and $L$, which are spacelike separated.
In region $R$ there are two alternative possible measurements, R1 and R2.
In region L there are two alternative possible measurements, L1 and L2.
Each local experiment has two alternative possible outcomes, 
labelled by + and --. The symbol $R1$ appearing in a logical statement stands 
for the statement ``Experiment R1 is chosen and performed in region R.'' 
The symbol $R1+$ stands for the statement that ``The outcome `+' of 
experiment R1 appears in region R.'' Analogous statements with other
variables have the analogous meanings.

There are, in the Hardy-type experimental set up, four pertinent predictions 
of quantum theory. They are expressed by the four logical statements:
$$
 (L2\wedge R2\wedge R2+)\Rightarrow (L2\wedge R2\wedge L2+). \eqno(3.1)
$$
$$
 (L2\wedge R1\wedge L2+)\Rightarrow (L2\wedge R1\wedge R1-). \eqno(3.2)
$$
$$
(L1\wedge R2\wedge L1-) \Rightarrow (L1\wedge R2\wedge R2+). \eqno(3.3)
$$
$$
 \neg [(L1\wedge R1\wedge L1-) \Rightarrow (R1-)]. \eqno(3.4)
$$
The symbol $\neg $ in front of the square brackets in (3.4) means that the 
statement in the square brackets is false.

The detectors are assumed to be 100\% efficienct, so that for each possible
world some outcome, either $+$ or $-$, will, according to quantum mechanics, 
appear in each of the two regions: for each (Ri, Lj), with $i$ and $j$ either
1 or 2, each world W in $\{Ri\}\cap\{Lj\}$ lies somewhere in 
$\{Ri+\}\cup\{Ri-\}$ and somewhere in $\{Lj+\}\cup\{Lj-\}$\hfill (3.5)

Each line of the following proof is a strict consequence of these
predictions of quantum mechanics, combined with the definition of LOC1
and the properties of the rudimentary logical symbols, plus the 
assumption LOC2, which is the assertion that line 6 of the proof follows
from line 5 by virtue of the demand that no influence can act backward in 
time in any frame. LOC2 is discussed in the next section.

\noindent {\bf Proof}:

1. $(L2\wedge R2\wedge L2+)
   \Rightarrow [R1\Box\rightarrow (L2\wedge R1\wedge L2+)].$ \hfill [LOC1c]

2. $(L2\wedge R2\wedge R2+) \Rightarrow (L2\wedge R2\wedge L2+).$ \hfill [3.1]

3. $(L2\wedge R1\wedge L2+)
   \Rightarrow (L2\wedge R1\wedge R1-)].$ \hfill [3.2]

4. $(L2\wedge R2\wedge R2+)
\Rightarrow [R1\Box\rightarrow (L2\wedge R1\wedge R1-)].$ \hfill [From 1, 
2, and 3]

5. $L2\Rightarrow [(R2\wedge R2+)\rightarrow (R1\Box
   \rightarrow R1\wedge R1-)].$ \hfill [LOC1e, (2.1)] 

6. $L1\Rightarrow [(R2\wedge R2+)\rightarrow (R1\Box
   \rightarrow R1\wedge R1-)].$ \hfill [LOC2]

7. $(L1\wedge R2) \Rightarrow [ R2+ \rightarrow (R1\Box
   \rightarrow R1\wedge R1-)].$ \hfill [LOGIC]

8. $(L1\wedge R2) \Rightarrow [ L1-\rightarrow R2+].$ \hfill [3.3]

9. $(L1\wedge R2) \Rightarrow [ L1-\rightarrow 
   (R1 \Box\rightarrow R1\wedge R1-)].$ \hfill [From 7 and 8.]

10. $(L1\wedge R2\wedge L1-)\Rightarrow  
   (R1 \Box\rightarrow R1\wedge R1-)$ \hfill (2.1)

11. $(L1\wedge R2)\Rightarrow 
   [R1 \Box\rightarrow (L1-\rightarrow R1\wedge R1-)].$ \hfill [LOC1d]
 
12. $L1\Rightarrow [R1 \rightarrow \neg (L1-\rightarrow R1\wedge R1-)].$ 
    \hfill [3.4]

13. $L1\Rightarrow [R1\Box\rightarrow\neg (L1-\rightarrow R1\wedge R1-)].$
       \hfill [(LOC1f)]

14. $(L1\wedge R2)\Rightarrow [R1\Box \rightarrow 
    \neg [L1- \rightarrow (R1\wedge R1-)]].$  \hfill [LOGIC.]
 
But the conjunction of $11$ and $14$ contradicts the assumption that the 
experimenters in regions  $R$ and  $L$ are free to choose which 
experiments they will perform. (See Appendix for details.) Thus the 
incompatibility of the assumptions of the proof is established.  

[Note that there is only one strict conditional [$\Rightarrow $] in each
line. In reference 10, some material conditionals standing to the right of
this strict conditional were mistakenly represented by the double arrow
$\Rightarrow $, rather than by $\rightarrow$. I thank Abner Shimomy and
Howard Stein [19] for alerting me this notational error.]

\vspace{.2in}
\noindent {\bf 4. The condition LOC 2.}
\vspace{.2in}

According to the most strict construal of the Copenhagen interpretation, 
the quantum formalism is merely a set of rules for making predictions 
pertaining to the appearance to human observers, under specified conditions, 
of events meeting certain specifications. Nevertheless, the normal idea among 
quantum physicists is that certain real events actually occur in association 
with measuring devices that are actually set in place. Dirac[14] speaks of 
``a choice on the part of `nature' '', and Heisenberg[15] says that 
``we may say that the transition from `possible' to `actual'  takes 
place as soon as the interaction of the object with the measuring device, 
and thereby with the rest of the world, has come into play.'' But there is a 
strong emphasis on the stipulation that these `choices on the part of 
nature' or `transitions from the possible to the actual' occur {\it only} 
if the measuring device is actually in place: {\it one cannot assume} 
that what `would have happened' in some experiment is well defined, even
theoretically,  unless that experiment is actually performed.

Although the quantum precepts unequivocally reject any {\it assumption} or 
{\it presumption} that, for an unperformed measurement, ``what would appear''
can be theoretically defined, these precepts do not absolutely rule out 
this possibility in {\it every} conceivable situation. Indeed, the basic 
intent here is to try to retain at least the notion that the outcome of a 
measurement that ``has already been completed'' can be assumed not to 
depend upon which measurement will be freely chosen and performed later.  
A  denial of this notion would seem to contradict Bohr's assertion, in his 
reply [16] to Einstein, Rosen, and Podolsky, that ``there is no question of 
a mechanical disturbance of the system under investigation''. For that 
debate centered on the possible dependence of properties of a system under 
investigation upon which  experiment is chosen and performed, say later,
on some other system. Yet any such nondependence of an earlier outcome on 
the later choice of which experiment is performed faraway would entail 
the theoretical existence of what `would have happened' earlier if the 
later choice had been different from what it actually turns out to be.

A logical framework for dealing in a rigorous way with this simplest putative 
locality condition, without violating any quantum precept, was described 
in section 2. The conclusions drawn that from that framework, alone, appear 
to be compatible with the predictions of quantum theory: I have found
no way to deduce any contradiction from that framework alone. This means that
the locality idea that no {\it outcome} can depend upon the far-away choice
apparently can be added to quantum theory without contradiction. This would
validate the apparent intent of Bohr's remark about no {\it mechanical
disturbance} [See D. Mermin [17].] 

This first theoretical constraint leads only to line 5 of the above proof.  
To go further let us consider that statement, line 5,  in a Lorentz frame 
in which region L is later than region R. Line 5 then asserts, given 
assumption LOC1 and truth of the statement that $L2$ is performed at the 
later time, the truth of the following statement: If under the condition 
that $R2$ is actually performed in the earlier region $R$ and that nature 
picks the outcome $R2+$, then if, instead, $R1$ had been freely chosen 
performed in region $R$, nature {\it would have picked} the outcome $R1-$.

This places a rigid constraint on nature's operation earlier in region $R$,
under the condition that $L2$ is performed later. Yet if that constraint
were to dissolve if $L1$ were to be chosen and performed later then there
would be {\it some sort} of influence of the later choice upon nature's
earlier operations in $R$.   

Of course, the assumption that $L2$ is performed in region $L$, is crucial 
to the {\it proof}, under condition LOC1, of the existence of the constraint 
in region $R$: one cannot {\it deduce} the nondependence of this constraint 
in $R$ on what is freely chosen in region $L$ merely from the premises that 
entailed the existence of this constraint. The assertion that there is no 
back action of this kind of the choice made in $L$ is thus an independent 
assumption.  I call it LOC2.   

\vspace{.2in} 
\noindent{\bf 5. Consistent Histories}
\vspace{.2in}

The basic question pertaining to the use of counterfactual reasoning to 
deduce a conflict between nonlocality and the predictions of quantum
theory is whether the counterfactual reasoning used in those proofs is 
itself compatible with the precepts of quantum theory.

One can, of course, just assert that any counterfactual reasoning that
leads to a conflict with locality is incompatible with quantum philosophy. 
But that fails to address the issue. The problem is that quantum theory is
manifestly nonlocal at the computational level: when one obtains 
information in one region about an entangled system, one is instructed to 
apply a projection operator that automatically reduces the state
also far away. This itself is no immediate problem, because an analogous
collapse is imposed for correlated systems in classical statistical  
mechanics, where there is no question of any real nonlocal effect: the
faraway change is understood to be simply a change in ``our knowledge''
of the faraway system. But there is a potential nonlocality problem in the
quantum case because in quantum theory the system itself is completely 
represented by the wave function that is supposed to represent 
``our knowledge''. This makes the faraway jump understandable, but only 
at the expense of allowing the complete representation of physical reality 
to violate locality. Hence it is not absolutely clear, a priori, that 
quantum theory necessarily is fully compatible with the putative locality 
condition that actions made in one place cannot influence faraway 
events.  

If one is to allow any sort of counterfactual reasoning at all, even in the
classical limit that emerges from quantum theory in the limit where Planck's 
constant is set to zero, one needs spell out what the general conditions are 
for valid counterfactual reasoning in a quantum context.

Robert Griffiths [18] has devised a set of conditions that ensure that the 
laws of classical logic can be applied consistently within a quantum context.
Recently [12] he has proposed a certain extension of that consistent 
histories framework. This extention is designed to allow the incorporation 
into quantum theory of certain counterfactual assertions that seem reasonable 
within the structure defined by the quantum precepts. The extended framework 
is explicitly quantum mechanical, and is guarenteed to produce no logical 
inconsistences. 

It is natural to inquire whether the nonlocality argument given above goes 
through within Griffiths' framework for incorporating counterfactual 
reasoning consistently into  quantum theory. 

Griffiths himself [12] applied his framework to a Hardy-type situation, and 
found no indication of any problem  with locality. But he noted that his 
framework was explicitly nonrelativistic, whereas my proof, for example,
depended on an explicitly assumed Lorentz invariance of the predictions
of the theory: this assumption is used to justify LOC2.

Actually, a direct application of Griffiths' framework to my proof is 
possible. The first phase of the Hardy-type argument that Griffiths examined 
is analogous to the first five lines of my proof. Griffiths examined this 
part of the argument from one particular viewpoint, and found that he could 
get nowhere at all. But he noted that other approaches are possible within his 
general framework.

Examination of the first part of my proof, using the 
``no-backward-in-time'' formulation appropriate for his 
nonrelativistic formulation, rather than the relativistic no-faster-than-light
formulation, reveals that one should use a frame in which the region $L$ is 
earlier in time than region $R$. [That is precisely the way that I formulated 
in ref. [10] the condition under which this first main conclusion was 
derived.] The natural set of projectors to use, to define the set of
consistent histories, are those that correspond to the set of alternative 
possible experimental arrangements, ordered in the way they are ordered in
nature. I shall use that natural set. 

The set of consistent histories is then uniquely defined.
It begins with
$$
[Hardy,\{L1,L2\},\{L1\& L1+,L1\& L1-,L2\& L2+, L2\& L2-\}], \eqno(5.1) 
$$
where the symbols $L1$, $L1+$, etc. now stand for the projection operators
corresponding to the `Yes' answers to the corresponding statements.
Then, at a later time, there is the set of 8 projectors
$$
\{L1\& L1\!+\& R1,L1\& L1\!+\& R2,L1\& L1\!-\& R1,L1\& L1\!- \& R2,
$$
$$  
 L2\& L2\!+\& R1,L2\& L2\!+\& R2,L2\& L2\!-\& R1,L2\& L2\!-\& R2\}. \eqno(5.2)
$$
Then, at a still later time, there is the set of 16 projectors
$$
\{L1\& L1\!+\& R1\& R1+, ...., L2\& L2\!-\& R2 \& R2-\}. \eqno(5.3)
$$

Line 5 of my above proof then follows directly from Griffiths' rules':   
if one starts in sector $L2$, and at $R2\wedge R2+$ and traces a path
back (i.e., back in time) to the place just before the choice between $R2$ 
and $R1$, and then traces this  path forward from that decision point along 
the $R1$ branch one finds that one must end up in $R1\wedge R1-$. The 
reason is that the Hardy condition that there be no state $L2-\wedge R2+$ 
forces the $L2\wedge R2+$ starting point to be $L2+\wedge R2+$. But the 
$R1/R2$ decision (``Pivot'') point --- on this path that starts 
at $L2+\wedge R2+$, and moves back in time to just before the $R1/R2$ 
decision --- is reached without passing through any $L$ decision points:
they all lie earlier. This means that when the path moves forward within 
the $R1$ sector it is still in the $L2+$ sector. But then the Hardy condition 
that there is no state $L2+\wedge R2+$ forces the returning path to end up 
in $R1-$, as specified in line 5. The conclusion, line 5, thus follows 
directly from Griffiths' rules.
 
The key question is whether one can now impose LOC2, specified by
line 6 of the proof, without generating a contradiction with the
other two predictions of quantum theory, (3.3) and (3.4).

Lines 6, and 7 of the proof both claim, in terms of Griffiths' formalism,
that if one starts at $L1\wedge R2\wedge R2+$ and then traces back to
the $R1/R2$ pivot point, and then moves forward along the $R1$ branch,
the one will end up in $R1-$, not $R1+$. Line 8 entails that if the
starting condition had been $L1\wedge R2\wedge L1-$ then, because there
is (in the Hardy state) no possible state 
$L1\wedge R2\wedge \L1- \wedge R2-$, one must be starting also from
$L1\wedge R2\wedge R2+$, as specified in lines 6 and 7. Thus one obtains
line 9 of the proof, in the form
$$
(L1\wedge R2\wedge L1-)\Rightarrow (R1\Box\rightarrow (R1\wedge R1-)).
\eqno(5.4)
$$
This asserts that starting from $L1\wedge R2\wedge L1-$ and tracing back 
to the $R1/ R2$ pivot point, and then forward along the $R1$ branch takes one 
to $R1\wedge R1-$.  But according to the  prediction of quantum 
theory (3.4) the probability for $R1-$ is not $100 \%$: the other possible 
outcome $R1+$ has a nonzero probablity to appear. Thus LOC2 leads to a 
contradiction {\it also} within Griffiths' framework for consistent 
counterfactual reasoning within quantum theory. 
 
Griffiths' claim is borne out: if one just stays within his formalism
no contradiction arises. For LOC2 does not arise within his formalism,
although the consequence of LOC1, line 5, does come out of his 
formalism. But one can say more: LOC2, if assumed, leads to a contradiction.
Thus the two methods of counterfactual reasoning within quantum theory 
both lead to the conclusion that an {\it outcome} that in some frame lies 
earlier than a later free choice can consistently be taken to be independent 
of that later choice, and that this entails line 5. But then assuming the 
Lorentz invariance of this first result, going to a  frame in which the 
time ordering of the two spacelike separated regions is reversed, and 
demanding there the absence of backward-in-time effects, entails that line 5 
should hold also with L1 in place of L2. But that condition leads, both in my
approach and by using Griffiths' prodedure, to a contradiction with two 
other predictions entailed by the Hardy state.

\vspace{.2in} 
\noindent{\bf 6. Conclusions}
\vspace{.2in}

It is possible to introduce into quantum theory, without generating
any apparent contradictions, the notion that what has already been
measured and recorded at times prior to a time T, as measured in any
given rest frame, does not depend upon which experiments are performed later.
This notion of the causal evolution of physical systems brings counterfactual 
reasoning into quantum theory in a way that is, I believe, completely 
compatible with the basic quantum principles: quantum physicists do not 
normally maintain that the outcomes they observe today could depend upon 
what other experimenters will decide to do tomorrow. Certainly Bohr never 
mentioned that idea in order to evade the challenges of Einstein. 

However, the acceptance of this weak locality property has a price!

The simplest way to state the situation is to note that what has been proved
is that the statement (R2+ implies that if, instead, R1 then R1-) is true 
if L2 is performed at the later time but is false if L1 is performed at the
later time. If we combine this with the orthodox idea that ``nature chooses''
a particular outcome in region R at the time of the occurrence 
(or observation) of the outcome in R of which ever measurement is performed
there, then nature's choice in the not actually realized case R1 would 
be logically forced always to be R1- in the case L2 is chosen later but 
sometimes be R2+ in case L1 is chosen later. The conditions that logically
entail this are that the predictions of the theory should hold whichever
conditions are set up, and that what nature chooses in R does not depend
on which experiment is chosen in L. 

These conditions are different from the hidden-variable conditions,
which entail immediately, quite independently of spacetime locations, that
in any given instance the outcomes of all four possible experiments
are simultaneously definable [8.9]. That hidden-variable assumption is 
manifestly at odds with quantum philosophy, whereas our conditions are 
formulated within the general framework of orthodox quantum thinking.

\newpage  
\begin{center}
{\bf Acknowledgements.}\\
It is a pleasure to thank P. Eberhard, J. Finkelstein, and G. Shapiro for 
helpful suggestions.
\vspace{.2in}

\vspace{.2in}
\noindent {\bf References}
\vspace{.2in}
\end{center}
1. Physics Today, December 1998, 9.

2. W. Tittle, J. Brendel, H. Zbinden, and N. Gisin, 
        Phys. Rev. Lett. {\bf 81}, 3563 (1998).

3. W. Tittle, J. Brendel, H. Zbinden, and N. Gisin, 
       quant-ph/9809025.
 
4. P.G. Kwiat, E. Waks, A. White, I. Appelbaum, and Philippe Eberhardt,
      quant-ph/9810003. 

5. G Weihs, T. Jennewein, C. Simon, H. Weinfurter, and A. Zeilinger,
       Phys. Rev, Lett. {\bf 81}, 5039 (1998).

6. J.S. Bell, Physics, {\bf 1}, 195 (1964);  
    J. Clauser and A. Shimony, Rep. Prog. Phys. {\bf 41}, 1881 (1978).

7. A. Einstein, N. Rosen, and B. Podolsky, Phys. Rev. {\bf 47}, 777 (1935). 

8. H. Stapp, Epistemological Letters, June 1978. (Assoc. F Gonseth,
          Case Postal 1081, Bienne Switzerland).

9. A Fine, Phys. Rev. Lett. {\bf 48}, 291 (1982).

10. H. Stapp, Amer. J. Phys. {\bf 65}, 300 (1997).
  
11. W Unruh, Phys. Rev. A. (To appear)

12. R. Griffiths, Phys. Rev. A. (Submitted)

13. L. Hardy, Phys. Rev. Lett. {\bf 71}, 1665 (1993); P. Kwiat, P.Eberhard
A. Steinberg, and R. Chiao, Phys. Rev. {\bf A49}, 3209 (1994).

14. Niels Bohr.  in A. Einstein, {\it Albert Einstein: Philosopher-Physicist}
ed, P.A. Schilpp, Tudor, New York, 1951.

15. Heisenberg, W. (1958b) {\it Physics and Philosophy} (New York: Harper and 
     Row).

16. N. Bohr, Phys. Rev. {\bf 48}, 696-702 (1935).

17. D. Mermin, Amer. J. Phys. {\bf 66} 920 (1998).

18. R. Griffiths, Phys. Rev. A {\bf 54} 2759 (1996).

19. A. Shimony and H. Stein, in {\it Space-Time, Quantum Entanglement, and
Critical Entanglement: Essays in Honor of John Stachel.} eds. A. Ashtekar, 
R.S Cohen, D. Howard. J. Renn, S. Sakar, and A. Shimony  
(Dordrecht: Kluwer Academic Publishers) 2000.

\newpage

\noindent {\bf APPENDIX: Set-theoretic proofs of key equations.}

For any statement $S$ expressed in terms of the rudimentary logical connections
let $\{W:S\}$ be the set of all (physically possible) worlds $W$ such that
statement $S$ is true at $W$ (i.e., $S$ is true in world W). Sometimes 
$\{W:S\}$ will be shortened to $\{S\}$.

The main set-theoretic equation is this: Suppose $A$ and $B$ are two
statements expressed in terms of the rudimentary logical connections.
Then $A\Rightarrow B$ is true if and only if 
the intersection of $\{A\}$ and $\{\neg B\}$ is void:
$$
[A\Rightarrow B] \equiv  [\{A\}\cap\{\neg B\}=\emptyset].  \eqno (A.1)
$$
Equivalently, $\{A\}$ is a subset of $\{B\}$:
$$
[A\Rightarrow B] \equiv  [\{A\} \subset \{B\}].  \eqno (A.2)
$$

Let $(S)_W$ mean that the statement $S$ is true at $W$. Then
$$
(A\rightarrow B)_W \equiv [(\neg A)_W \mbox{ or } (B)_W].     \eqno (A.3) 
$$
Equivalently,
$$
\{A\rightarrow B\}\equiv [\{\neg A\}\cup \{B\}]. \eqno (A.4)
$$
\begin{center}
 {\bf Proof of (2.1)}
\end{center}

Equation (2.1) reads:
$$
[A\Rightarrow (B\rightarrow C)]\equiv [(A\cap B)\Rightarrow C]. \eqno (A.5)
$$
This is equivalent to
$$
[\{A\}\cap\{\neg (B\rightarrow C)\}=\emptyset]\\
\equiv [\{A\cap B\}\cap \{\neg C\}=\emptyset].    \eqno (A.6)
$$
But $\{\neg (B\rightarrow C)\}$ is the complement of 
$\{B\rightarrow C\}$. Using (A.4), and the fact that the complement
of $\{\neg B\} \cup \{C\}$ is $\{B\} \cap \{\neg C\}$, one obtains the
needed result.  

\begin{center}
{\bf Set-theoretic forms of $C\Box\rightarrow D.$}
\end{center}

According to the definition at (2.5), 
$$
(C\Box\rightarrow D)_W \equiv  
$$
$$
[(D)_V \mbox{ for every V such that } (C)_V \mbox{ and } (V=W 
\mbox{ outside }V^+(C/W))]. \eqno (A.7)
$$

Equivalently, 
$$
\{W: C\Box\rightarrow D\} \equiv
$$
$$
\{W:\{\{V:V=W\mbox{ outside }V^+(C/W)\}\cap\{V:C\}\}\subset\{V:D\}\}.
\eqno (A.8)
$$
 
\begin{center}
{\bf Proof of (LOC1c).}
\end{center}

LOC1c reads 
$$
(L2\wedge R2\wedge L2+)\Rightarrow [R1\Box\rightarrow 
(L2\wedge R1 \wedge L2+)]. \eqno (A.9)
$$
To prove (A.9) it is sufficient, by virtue of (A.2) and (A.8),
to show that
$$
\{\{W:L2\}\cap\{W:R2\}\cap\{W:L2+\}\}\subset
$$
$$
\{W:\{\{V:V=W\mbox{ outside }V^+(R1/W)\}\cap\{V:R1\}\} 
$$
$$
\subset\{\{V:L2\}\cap\{V:R1\}\cap\{V:L2+\}\}\}. 
\eqno (A.10)
$$
Let W be any world satisfying the conditions on the LHS of (A.10). 
Then $V^+(R1/W)$ is $V^+(R)$, the forward light cone from the region R. 
But for any W in $\{W: L2\}\cap\{W:L2+\}$ the set 
$\{V:V=W\mbox{ outside } V^+(R)\}\subset \{\{V:L2\}\cap\{V:L2+\}\}$.
This entails the truth of (A.10).

\begin{center}
{\bf Proof of (LOC1d)}.
\end{center}

LOC1d asserts that if condition B is localized in a region that lies
outside $V^+(C)$, the forward light cone from the region in which
condition C is localized, then
$$
[(A\wedge B)\Rightarrow (C\Box\rightarrow D]\equiv
[A\Rightarrow (C\Box\rightarrow (B\rightarrow D))].
\eqno (A.11)
$$
Equation (A.11) is, by virtue of (A.2) and (A.4), equivalent to the 
assertion that (A.12) and (A.13) are equivalent:
$$
\{\{W:A\}\cap\{W:B\}\}\subset
$$
$$
\{W:\{\{V:V=W\mbox{ outside }V^+(C/W)\}\cap\{V:C\}\}
\subset\{V:D\}\}   \eqno (A.12)
$$
and
$$
\{W:A\}\subset
$$
$$
\{W:\{\{V:V=W\mbox{ outside }V^+(C/W)\}\cap\{V:C\}\}
\subset\{V:D\}\cup\{V:\neg B\}\}.   \eqno (A.13)
$$
Because condition B is localized outside of $V^+(C)$ it
is localized also outside of $V^+(C/W)$ for all W.
But then (A.12) and (A.13) are both equivalent to
$$
\{\{W:A\}\cap\{W:B\}\}\subset
$$
$$
\{W:\{\{V:V=W\mbox{ outside }V^+(C/W)\}\cap\{V:C\}\}
\subset\{\{V:D\}\cup\{V:\neg B\}\}\}.   \eqno (A.14)
$$
Equation (A.12) is equivalent to (A.14) because the
condition B imposed on W=V in both equations renders 
the condition $\cup\{V:\neg B\}$ inoperative.
Equation (A.13) is equivalent to (A.14) because
the condition $\cup\{V:\neg B\}$ imposed on V=W in both 
of these equations renders the condition B on W immaterial: 
relaxing in (A.14) the condition B on the W=V enlarges
throughout the equation the set of W=V by the set at which 
B is false.

\begin{center}
{\bf Proof of LOC1e}
\end{center}

The LHS of LOC1e is, by virtue of (A.2) and (A.8), equivalent 
to
$$
\{\{W:A\}\cap\{W:B\}\}\subset
$$
$$
\{W:\{\{V:V=W\mbox{ outside }V^+(C/W)\}\cap\{V:C\}\}
\subset\{\{V:D\}\cap\{V:B\}\}\}.   \eqno (A.15)
$$
The condition on B entails, as before, that condition B is 
localized outside $V^+(C/W)$ for all W.  Hence the condition
B on W entails also the condition B on V.  Hence the final
condition $\cap\{B\}$ is automatic: it is entailed by the
condition B on W, and adds no extra condition. This is
precisely what LOC1e asserts.

\begin{center}
{\bf Proof of LOC1f}
\end{center}

LOC1f asserts that, under the same condition that B lies outside $V^+(C)$,
$$
[B\Rightarrow (C\rightarrow D)]\Rightarrow[B\Rightarrow
(C\Box\rightarrow D)]. \eqno (A.16)
$$
By virtue of (A.2) and (A.4) the LHS is equivalent to
$$
\{W:B\}\subset\{\{W:D\}\cup\{W:\neg C\}\}. \eqno(A.17)
$$
By virtue of (A.2) and (A.8) the RHS is equivalent to
$$
\{W:B\}\subset 
$$
$$
\{W:\{\{V:V=W\mbox{ outside }V^+(C/W)\}\cap\{V:C\}\}\subset\{V:D\}\}.
\eqno (A.18)
$$
As before, condition B is localized outside $V^+(C/W)$, and hence
V=W satisfies B. Thus, (A.17) entails that V lies in
$\{V:D\}\cup\{V:\neg C\}$. But then if V lies also in 
$\{V:C\}$, as specified in (A.18), then V lies in
$\{D\}$, as demanded by (A.18). Thus the condition
(A.17) entail the condition (A.18), and hence, by virtue
of (A.2), equation (A.16) holds.

\begin{center}
{\bf Proof of Line 12}
\end{center}

Line 12 of the proof is, by virtue of (2.1), equivalent to
$$
(L1\wedge R1)\Rightarrow\neg (L1-\rightarrow R1\wedge R1-).
\eqno (A.19)
$$
By virtue of (A.1) and (A.4) this is equivalent to
$$
\{L1\}\cap\{R1\}\cap\{\{R1-\}\cup\{\neg L1-\}\}=
\emptyset. 
$$
This implies, by virtue of (3.5), 
$$
\{L1\}\cap\{R1\}\cap\{\{\neg R1-\}\cap\{ L1-\}\}\neq
\emptyset. \eqno(A.20)
$$
But equation (3.4) is, by virtue of (A.1), equivalent to
$$
\{L1\}\cap\{R1\}\cap\{L1-\}\cap\{\neg R1-\} \neq
\emptyset. \eqno(A.21)
$$

\begin{center}
{\bf Proof of the incompatibility of lines 11 and 14 of the proof.}
\end{center}

According to (A.2) and (A.8), line 11 is equivalent to
$$
\{\{W:L1\}\cap\{W:R2\}\}\subset
$$
$$
\{W:\{\{V:V=W\mbox{ outside }V^+(R1/W)\}\cap\{V:R1\}\}
$$
$$
\cap \{\neg R1-\}\cap\{L1-\}=\emptyset\}.   \eqno (A.22)
$$
On the other hand, line 14 asserts that
$$
\{\{W:L1\}\cap\{W:R2\}\}\subset
$$
$$
\{W:\{\{V:V=W\mbox{ outside }V^+(R1/W)\}\cap\{V:R1\}\}
$$
$$
\cap\{\{R1-\}\}\cup\{\neg L1-\}\}\}=\emptyset\}.   
\eqno (A.23)
$$
Condition (A.22) entails that if W=V lies in $\{L1\wedge L1- \}$
and V lies in $\{R1\}$ then V cannot lie in $\{\neg R1-\}$. But equation 
(A.23) entails that if W=V lies in $\{ L1\wedge L1-\}$ and V lies 
$\{R1\}$ then V cannot lie in $\{R1-\}$. But, by virtue on (3.5), there 
are W such that W lies in $\{L1\wedge L1-\}$; and any V in $\{R1\}$
must lie in either $\{R1-\}$ or  $\{\neg R1-\}$. Thus lines 11 and 14 are
incompatible. 

\end{document}